\documentstyle[12pt]{article}
\begin{document}
\baselineskip 24pt
\newcommand{\loo}{\,\raisebox{-.5ex}{$\stackrel{<}{\scriptstyle\sim}$}\,}
\vspace{0.5 truecm}

\begin{center}
{\Large \bf  Partitioning Composite Finite Systems}
\end{center}
\vspace{0.5 truecm}
\begin{center}
{\large A.S. Botvina$^{1,2,3}$, A.D. Jackson$^4$,
and I.N. Mishustin$^{4,5,6}$}
\end{center}
\vspace{0.3 truecm}
\begin{center}
{\it 
$^{1}$GANIL (CEA-DSM/CNRS-IN2P3), B.P.5027, F-14076 Caen Cedex 5, France\\
$^{2}$Dipartimento di Fisica and INFN, 40126 Bologna, Italy\\
$^{3}$Institute for Nuclear Research, Russian Academy of Science, 
 117312 Moscow, Russia\\
$^{4}$Niels Bohr Institute, DK-2100 Copenhagen {\O}, Denmark\\
$^{5}$Kurchatov Institute, Russian Research Center, 123182 Moscow, Russia\\
$^{6}$Institute for Theoretical Physics, J.-W. Goethe University, D-60054 
Frankfurt am Main, Germany}
\end{center}

\vspace{0.5 truecm}
\begin{abstract} 
We compare different analytical and numerical methods  
for studying the partitions of a finite system into fragments. 
We propose a new numerical method of exploring the partition space by 
generating the Markov chains of partitions based on  the Metropolis algorithm. 
The advantages of the new method for the problems where partitions are 
sampled with non-trivial weights are demonstrated. 
\end{abstract}
\vspace{0.3 truecm}

\hspace{0.5 truecm}PACS numbers: 25.70.Pq, 02.70.Lq .
\vspace{0.5 truecm}

Many fields of physics deal with the common phenomenon that, under 
appropriate conditions, a compound system can disintegrate into constituents. 
Let us consider an isolated system composed of $A_0$ identical particles
(we call them nucleons) which are kept together by some attractive forces. 
If sufficient energy is put into the system, it will disintegrate into 
fragments.  These fragments can either be individual nucleons or bound 
clusters of several nucleons.  Examples of such processes abound in 
condensed matter physics, nuclear physics, and astrophysics.  In order to 
provide a microscopic description of such processes, one must sort out 
possible partitions of the system and compare their probabilities.  At the 
first step, it is necessary to develop methods of generating and sampling the
partitions. The aim of this paper is to propose a new and efficient method 
of doing this. 

The obvious way to proceed is simply to construct all partitions directly and 
calculate the characteristics of interest.  Unfortunately, this approach is 
possible only for small $A_0$ because the total number of partitions, $P(A_0)$,
grows rapidly with $A_0$. For instance, $P(100)=190569292$ while 
$P(200)=3972999029388$. 
Even if one needs only perform a few non-trivial operations for each 
partition, this task becomes intractable for $A_0>100$.  We shall, however, 
reserve this direct method for checking the more practical methods 
presented below.

First, we address an analytical approach to dealing with the Euler's 
partitioning problem. It is based on the Generating Function (GF) formalism 
\cite{And}.  This approach can be applied successfully for calculating average 
characteristics of partitions. We characterize each partition $f$ by the 
multiplicities $\{N_A\}$ of fragments with different nucleon numbers $A$, 
$1\le A\le A_0$. Then, the conservation of the total nucleon number for each $f$
is expressed as: 
\begin{equation} \label{eq:na}
\sum_{A=1}^{A_0}N_{A}^{(f)}A=A_{0}~.
\end{equation}
Evidently, the total fragment multiplicity $M$ in the channel $f$ is 
\begin{equation}
M_{f}=\sum_{A=1}^{A_0}N_{A}^{(f)}~.
\end{equation}
Following a well-established method in mathematical literature \cite{And}, 
we introduce an unconstrained generating function $Z(x)$:
\begin{equation} \label{eq:zx}
Z(x)=\sum_{N_{A}=0}^{\infty}\prod_{A=1}^{\infty}
\left(c_{A}x^{A}\right)^{N_{A}}=\prod_{A=1}^{\infty}
\frac{1}{1-c_{A}x^{A}}~,
\end{equation}
where the $c_A$ are arbitrary numbers which can later be taken as 
$c_A$=1. Here $x$ can be considered as a Lagrange multiplier.
Now we can calculate the total number of partitions, $P(A_0)$, by simply 
expanding eq.(\ref{eq:zx}) and counting the coefficient of $x^{A_0}$. 
The results for large $A_0$ or $x\to 1$ are well approximated by famous 
Hardy-Ramanujan formula:
\begin{equation}
P(A_0)=\frac{1}{\sqrt{48}A_0}\cdot \exp\left(\pi\sqrt{\frac{2A_0}{3}}\right)
+O\left(\left[\exp\left(\pi\sqrt{\frac{2A_0}{3}}\right)\right]^{1/2}\right).
\end{equation}
One can use this generating function to calculate approximately the average 
multiplicities of fragments $\langle N_A \rangle$ over all partitions. 
This is done by replacing the exact constraint of eq. (\ref{eq:na}) by an 
approximate one: 
\begin{equation} 
\sum_{A=1}^{\infty}\langle N_{A} \rangle A=A_{0} \, ,
\end{equation}
i.e. the constraint is fulfilled on average only. Then one obtains 
\begin{equation} \label{eq:a0}
A_0=x\frac{\partial \ln\left(Z(x)\right)}{\partial x}=
\sum_{A=1}^{\infty}\frac{Ax^{A}}{1-x^{A}}~,  
\end{equation}
where we have set the $c_A=1$.  This equation must be solved to 
determine $x$. A very good approximation to the solution at large $A_0$ is 
\begin{equation}
x=\exp\left(-\pi\sqrt{\frac{1}{6A_0}}+\frac{1}{4A_0}\right).
\end{equation}
Now, the mean multiplicities of fragments can be calculated as:
\begin{equation} \label{eq:nas}
\langle N_A \rangle =c_{A}\frac{\partial \ln\left(Z(x)\right)}{\partial c_{A}}=
\frac{x^{A}}{1-x^{A}}. 
\end{equation}
The result is shown in fig.~1 (top panel) in comparison with the results 
of the direct method in which all the partitions are included in the 
calculation.  It is seen that the agreement is good except for a slight 
discrepancy at large $A$ which indicates an expected finite size effect. 
Indeed, eq. (\ref{eq:nas}) gives small but finite $\langle N_A\rangle$ even 
for $A>A_0$ when the exact calculation gives strictly zero. 
The average multiplicity of all fragments can be calculated as 
$\langle M \rangle = \sum_{A} \langle N_A \rangle$ and  
is well approximated by the expression
\begin{equation} \label{mult}
\langle M \rangle =\frac{1}{\pi}\sqrt{\frac{3A_0}{2}}
\ln\left(\frac{6A_{0}}{b\pi^2}\right), 
\end{equation}
with $b=$0.315087. For example, for 
$A_0$=100 it gives us $\langle M \rangle =21.32$ while the exact value 
obtained with the direct method is 21.75.

More generally, it is useful to consider the situation in which partitions 
are biased with certain weights.  In statistical theory, for example, 
identical fragments are counted in a partition sum with a 
factorial weight 
$1/N_{A}!$.  The weight of a partition is then $W_f=1/\prod_{A}N_{A}!$. 
In this case, the corresponding generating function can be written as:
\begin{equation} \label{eq:zw}
Z(x)=\sum_{N_{A}=0}^{\infty}\prod_{A=1}^{\infty}
\frac{\left(c_{A}x^{A}\right)^{N_{A}}}{N_{A}!}=\prod_{A=1}^{\infty}
\exp\left(c_{A}x^{A}\right).
\end{equation}
This form is similar to the grand canonical partition sum if one identifies $x$
with the fugacity and $c_A$'s with the internal partition sums of individual
fragments \cite{bond95}. Now instead of eqs.(\ref{eq:a0}) and (\ref{eq:nas}) 
one easily obtains (after substituting $c_A$=1):
\begin{equation} \label{eq:a0s}
A_0=\sum_{A=1}^{\infty}Ax^{A},~~\langle N_A \rangle =x^{A} \ . 
\end{equation}
For $A_0 \to \infty$ one finds the approximate expressions  
$x=\exp(-1/\sqrt{A_0})$ and $\langle M \rangle = \sqrt{A_0}$. These 
results are shown in fig.~1 (bottom panel).  The mean multiplicity 
$\langle M \rangle =10$ for the case $A_0 = 100$ is in good agreement 
with the exact value of 9.77 obtained by direct calculation. 

For the two simple examples considered above one can calculate also the 
multiplicity distributions of individual fragments. It is clear from
the structure of the generating functions, eqs. (\ref{eq:zx}) and
(\ref{eq:zw}), that the distribution is exponential in the first case and
Poissonian in the second case. The normalized multiplicity distributions 
are respectively,
\begin{equation} \label{multdis}
P_1(N_A)=\frac{1}{1+\langle N_A\rangle}\left(\frac{\langle N_A\rangle}{1+\langle
N_A\rangle}\right)^{N_A}~~~,~~P_2(N_A)= 
\exp{(-\langle N_A\rangle)}\frac{\langle N_A\rangle^{N_A}}{N_A!}.
\end{equation}  
As seen in fig. 3, the exact results are reproduced by these distributions 
with high accuracy.

In practice, however, direct accounting for all partitions can only be done 
for $A_0\loo 100$. If the weight factors are complicated, it can also be hard
to find an analytical solution. Multiplicity distributions and correlations, 
which are of considerable physical interest, are particularly difficult to 
obtain\footnote{In this respect an interesting development of an analytical 
method was recently made in ref. \cite{prat}.}. There is thus a need for 
another method, presumably based on the generation of individual partitions. 
Obviously, it must be efficient enough to permit computer simulation within a
reasonable time. 

A first attempt to develop such a method was made in 
refs.\,\cite{bond85} by introducing a bias function $b(A_0,M) = 
P(A_0,M)/P(A_0)$, where $P(A_0,M)$ is the total number of partitions with 
exactly $M$ fragments.  It can be calculated using the recursion relation 
\cite{And,bond85} 
\begin{equation} \label{eq:pam}
P(A_{0},M) \;=\; P(A_{0}-M,M)\; +\; P(A_{0}-1,M-1) \ .
\end{equation}
As before, the total number of partitions is
\begin{equation}
P(A_{0})=\sum_{M} P(A_{0},M).
\end{equation}
This bias function is used to generate a sample of partitions by the Monte 
Carlo method.  First, $M$ is selected randomly with a probability given by 
the bias function, $b(A_0,M)$.  Then, a random partition with selected 
multiplicity is generated as described in ref. \cite{bond95}.
We shall refer to this method as Biased Random Generation (BRG). 
Another Monte Carlo method of generating partition samples using a bias 
function obtained with a Laplace transformation is described in 
ref. \cite{gross90}. 

Figs.~1 and 2 (top panel) show 
how well the BRG  method works in the case when all partitions have equal 
weighs.  The results are presented for $A_0$=100 and summarize the outcome 
of 10$^5$ randomly generated partitions.  By construction, this method is 
guaranteed to give the correct multiplicity distribution as shown in fig.~2 
(top panel).  It is less trivial that it reproduces correctly also the mean 
multiplicities of individual fragments as well as other 
distributions.  Unfortunately, the BRG method has a serious drawback: It 
produces correct results only for the case in which the weights of partitions 
are equal.  This is not surprising given that eq.\,(\ref{eq:pam}) was 
obtained under this assumption.  When we introduce nontrivial weight factors, 
for instance relative factorial weights $W=1/\prod_{A}N_{A}!$ for 
partitions with fixed $M$, the method fails. This is clearly seen in the 
bottom panel of fig.~1 for mean fragment multiplicities.
In the case of nontrivial partition weights the calculation of a bias
function might be even more difficult than the calculation of a corresponding
generating function.

Here, we propose a new method of the partition sampling which is designed 
especially for computer simulations.  The idea is to generate a Markov chain 
by moving from one partition to another by minimal steps, i.e., by demanding 
that neighboring partitions differ by the state of one nucleon only.  We 
shall refer to this method of generating partition samples as Markov Chain 
Generation (MCG).  The procedure allows the following moves: 
(a) to transfer a nucleon from 
one fragment to another, (b) to make a nucleon free, or (c) to attach a free 
nucleon to a fragment.  In addition, one must ensure that each new partition 
is different from the previous one, since fragments 
with the same $A$ are to be regarded as indistinguishable.  

As well known, any sampling procedure of this kind must satisfy the detailed 
balance requirement. This can be achieved by applying the famous Metropolis 
algorithm \cite{metrop1}, where  a chain of partitions is generated by 
performing subsequent moves in the partition space biased by the partition 
probabilities $W$ (weight factors).
As shown elsewhere (e.g., ref.\,\cite{metrop2}), this method provides a 
correct description of the complete partition space for any specified weight 
factors $W$. In the MCG the number of all possible moves is limited and easily 
countable for any partition. By generating a new partition we account for the 
probability of all possible moves, and thus we avoid the bias function 
problem. Detailed balance is guaranteed by 
application of the Metropolis algorithm. 

The numerical procedure is implemented in the following way:\\
\noindent Step I: For a given partition with $M$ fragments of mass numbers 
$A_i$ ($i=1,...,M$), enumerate all fragments in the order of decreasing mass 
so that $A_1 \geq A_2 \geq ... \geq A_M$.  This order is to be strictly 
maintained; any move violating this ordering is rejected.  In this manner, 
we ensure that each move gives a genuinely new partition.\\
\noindent Step II:  Select at random the fragment $i$ that looses a nucleon 
and the fragment $j$ ($j=1,...,M+1; j\neq i$) that accepts it. 
(The case $j=M+1$ corresponds to making the nucleon free.)  Check this move 
against the ordering requirement of Step I.  If the order is violated, 
repeat the determination of $i$ and $j$. \\ 
\noindent Step III: Calculate the weight of a new partition, $W_{\rm new}$, 
and compare it with the weight of the previous one, $W_{\rm old}$.  
A new partition is added to the ensemble if $W_{\rm new} \geq W_{\rm old}$. 
If $W_{\rm new} <W_{\rm old}$, a new partition is added with probability 
$W_{\rm new}/W_{\rm old}$.  Otherwise, the old 
partition is taken as the new one and a new move is undertaken. \\
\noindent Step IV: Calculate the characteristics of interest by taking all 
partitions from the chain. The chain is truncated when these 
characteristics are saturated.

We stress that, contrary to the GF and BRG 
methods discussed above, the MCG method is a purely numerical procedure which 
requires nothing more than random number generation.  This provides a welcome 
degree of universality which is missing in other methods. For example, 
similar to the direct calculation, our method can be applied in case of any 
partition weight, as well as it 
can be easily generalized for other partition spaces, e.g., when 
fragments are characterized by two numbers (such as mass $A$ and charge $Z$) 
instead of one \cite{future}. 

The initialization problem, i.e., the question of which partition should be 
taken as a seed, does not appear to be important for the MCG 
method.  The system with $A_0$=100 loses all memory of the initial 
partition after approximately $10^4$ moves.  In order to obtain a 
representative partition sample, one should just discard these initial
partitions from the ensemble.  This is verified for several cases when 
partition weights vary smoothly with fragment mass and the number of 
fragments. In other cases, the number of initial moves may increase.  
This problem must be analyzed in each particular case.

We have checked the MCG method in a number of ways. The results are presented 
in figs.~1-4 for two cases: first, when all partitions have equal weights (top
panels) and second, when  partitions with identical fragments are 
suppressed by  the factorial weights
$W_f=1/\prod_{A}N_A!$ (bottom panels). They show the mean fragment 
multiplicity as a function of $A$ (the mass distribution), the distribution of 
total fragment multiplicity, and a very specific characteristic, i.e., the 
distribution of multiplicities of particular fragments ($A=1$, $A=4$ and 
$A \geq 10$) taken over all partitions.  
The results of the exact direct method and of Markov chain generation are 
in remarkably good agreement.  It should be stressed that for $A_0=100$ all 
$1.9 \times 10^8$ partitions are included in the direct method 
while only 10$^5$ partitions can be taken from the chain 
to explore the entire partition space with the MCG method.  
Small discrepancies in the tails of the distributions are seemingly related to
a limited sample size and numerical precision. 
However they are not important in practice because of their very 
small relative weight in the chain. 
For smaller systems (e.g. $A_0$=20) the agreement is also good. 
For larger systems (e.g., $A_0$=1000), where the direct method is intractable, 
comparisons were made with the analytical GF method.  
As demonstrated in fig.~4, the agreement is quite good, 
apart of a small discrepancy in the tails. One should bear in mind, however, 
that the GF method slightly overestimates the exact result (see fig.~1). 
We emphasize that the same high quality agreement between the direct 
and MCG methods is achieved in both considered cases which differ significantly
by the weight factors. 
Calculations have been made for partitioning with other weights, 
and similar agreement has been found.  Therefore, we believe that 
the MCG method described here offers a simple and correct numerical solution 
to the partition sampling problem.

In conclusion, we have analyzed several methods for calculating 
characteristics of the partition space of a finite composite system. 
We have developed a new numerical method, the Markov Chain Generation, 
which is flexible and efficient in practical calculations 
with complicated partition weights.  We see a variety of 
applications of this method in different fields dealing with finite-size 
objects, from atomic nuclei to molecular clusters 
and astrophysical objects.  We believe that this 
method will be very useful for studying the thermodynamics of finite systems
\cite{bond95,prat,gross90,future}. 

The authors thank J.P. Bondorf for fruitful discussions. 
A.S.B. thanks the INFN, Italy (Bologna section), and I.N.M. thanks the Niels 
Bohr Institute, Copenhagen University, 
for the kind hospitality and financial support. This work was supported in 
part by the Humboldt Foundation, Germany.

\newpage

{\large \bf  {Figure captions}}\\

{\bf Fig. 1.} {Average multiplicities $\langle N_A\rangle$ of fragments 
with mass number $A$ for  the system with total mass 
$A_0=100$. Solid lines: direct calculation taking into account all 
partitions, dashed lines: numerical Markov chain generation of 
partitions, dot-dashed lines: analytical calculations by the generating 
function method, dotted lines: biased random generation. 
Top panel: for partitions with equal weights,
bottom panel: for partitions with the factorial weights $1/\prod_{A}N_{A}!$.}

\vspace{0.5 cm}

{\bf Fig. 2.} {Distribution of total fragment multiplicity $M$ 
for the system $A_0=100$. Notations are the same as in fig.~1.} 

\vspace{0.5 cm}

{\bf Fig. 3.} {Multiplicity distributions of fragments with $A=1$, $A=4$ and
$A\geq 10$ for the system $A_0=100$. Notations are the same as in fig.~1.}

\vspace{0.5 cm}

{\bf Fig. 4.} {Comparison of fragment mass distributions for $A_0$=20 and 1000 
calculated by the analytical and Markov chain generation methods. Top and 
bottom panels show calculations for two different weighting factors as above.}

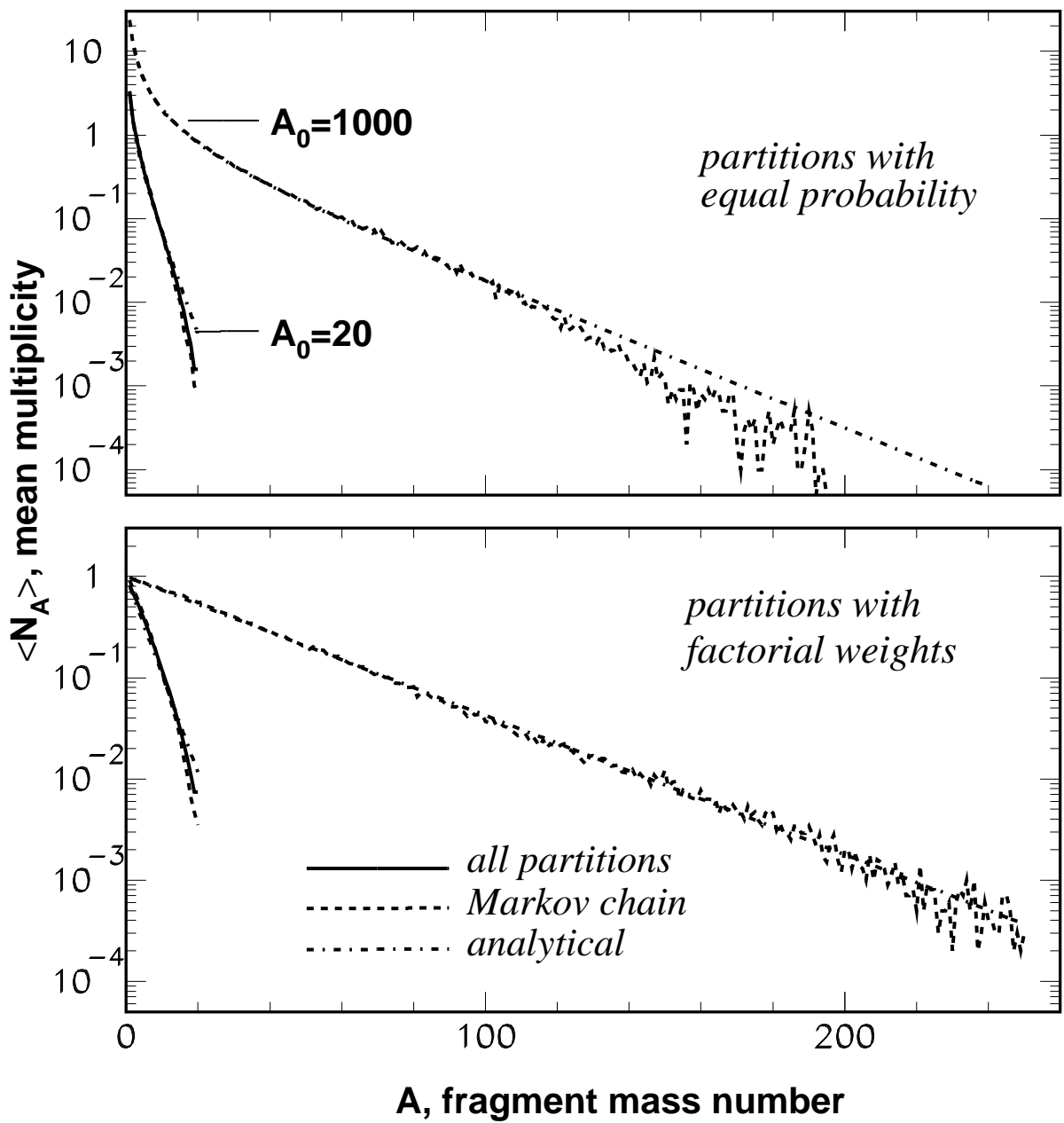
\end{document}